\documentclass[usegraphicx,usenatbib]{mn2e}
\usepackage{amssymb}
\usepackage[dvips]{graphicx}
\usepackage{longtable}
\usepackage{epstopdf}
\usepackage {graphicx}

\def\Teff{$T_{\rm eff}$}
\def\logg{$\log\,g$}

\def\Vt{V${\rm t}$}
\def\Vr{V${\rm r}$}
\def\Vbary{V$_{\rm bary}$}
\def\VLSR{V$_{\rm LSR}$}
\def\kms{${\rm km\,s^{-1}}$}
\def\Fe{$\mathrm{[Fe/H]}$}
\def\Rg{$R_{G}$}

\title[Abundances in the center of the Galaxy]
{Elemental abundances in the center of the Galactic Nuclear Disc}

\author[V.V. Kovtyukh et al.]
{V.~V. Kovtyukh$^{1,2}$\thanks{E-mail: vkovtyukh@ukr.net},
S.~M. Andrievsky$^{1,2,3}$,
R.~P. Martin$^{4}$\thanks{Visiting astronomer at the Infrared Telescope Facility, which is
operated by the University of Hawaii under contract NNH14CK55B with the
National Aeronautics and Space Administration.},
S.~A. Korotin$^{5}$,
\newauthor
J.~R.~D. Lepine$^{6}$,
W.~J.~Maciel$^{6}$,
L.~E. Keir$^{1}$,
E.~A. Panko$^{7}$
\\
$^1$Astronomical Observatory, Odessa National University, Shevchenko Park, 65014 Odessa, Ukraine\\
$^2$Isaac Newton Institute of Chile, Odessa Branch, Shevchenko Park, 65014 Odessa, Ukraine\\
$^3$GEPI, Observatoire de Paris-Meudon, CNRS, Universite Paris Diderot, 92125
Meudon Cedex, France\\
$^4$Department of Physics and Astronomy, University of Hawai'i at Hilo, Hilo,
HI, 96720, USA\\
$^5$Crimean Astrophysical Observatory, Nauchny 298409, Republic of Crimea\\
$^6$Instituto de Astronomia, Geof\'{\i}sica e Ci\^encias Atmosf\'ericas da
Universidade de S\~ao Paulo,\\ Cidade Universit\'aria, CEP: 05508-900, S\~ao Paulo, SP, Brazil\\
$^7$Department of theoretical physics and astronomy, Odessa National University,
42 Pasteur Str., Odessa, Ukraine\\
}

\begin{document} 

\date{Accepted 2019 August 14. Received 2019 August 1; in original form 2019 April 9}

\pagerange{\pageref{firstpage}--\pageref{lastpage}} \pubyear{2019}

\maketitle

\label{firstpage}

\begin{abstract}

We have made the first attempt to derive the chemical properties of the 
Galactic disc at its very central part using high-resolution infrared
spectroscopic observations of four classical Cepheids. Those stars are
located at Galactocentric distances smaller than 1 kpc. 
All investigated stars show near-to-solar elemental abundances. 
By combining these new data with our previous studies, this result suggests
that the radial distribution of iron abundance on a logarithmic scale gradually increases 
from the outskirts of the Galactic disc to Galactocentric distances of about 2--4 kpc, 
reaching there a maximal value of about +0.4 dex, and then declines sharply to about 
the solar value at the Galactic Center.

\end{abstract}


\section{Introduction}

In our two previous papers on the chemical properties of the central part in the 
Galactic thin disc \citep[see][]{Mar2015, And2016}, it has been suggested that there 
is a plateau in metallicity distribution within approximately 2$-$4  kpc from the 
Galactic Center. The maximum metallicity here reaches about [Fe/H] = +0.4 dex.
This value is attained following the more or less radial monotonic
increase of the metallicity from the outer parts of the Galactic disc
(16 kpc) to the Galactic Center, with a slope in the global [Fe/H]
abundance gradient of about --0.055 dex kpc$^{-1}$.

At the same time, several studies \citep[see compilation in][]{Mar2015}
show that the chemical properties at the very Galactic Center are almost
the same as those at the Galactocentric distance of the Sun.
Thus, within the range of 0 to approximately 3 kpc, the metallicity gradient even appears to have a positive slope. 

It should be noted, however, that the central part of the Galactic thin disc is
poorly sampled \citep[see Figs. from][]{And2016}. Additional observations are 
urgently needed to confirm (or disprove) the existence of the metallicity plateau 
within the very inner disc (which was suspected, in particular, by  \citealt{And2016}) and, also 
to determine its characteristics (slope, level) if present. 

Of particular interest is the determination of the metallicity in the center of the
Nuclear Disc. Several studies have been devoted to this topic. In \citeyear{Carr2000}, Carr et al. studied 
the infrared spectra of M2 IRS, a supergiant which is less than 10 million years old and 
located near the Galactic Center. They found a solar metallicity for this 
star (within the measurement errors). Moreover one M supergiant RV 5-7,
located at a Galactocentric 
distance of less than 30 pc, was studied by \cite{Ram2000}. These authors also 
found a solar metallicity value for that star. \cite{Cunha2007} also analyzed spectra of several 
cool supergiants including IRS 7 and VR 5-7. They derived 
masses of 22 and 14 solar masses for those stars, respectively, indicating  
relatively young ages. The mean [Fe/H] value derived for the sample of M supergiants 
was +0.14 dex. Similarly, \cite{Ryde2015} determined the metallicity in M giant stars, and
found a mean [Fe/H] = +0.11 dex. 

Summarizing, it can be stated that young M supergiants and older M giants 
located in the Galaxy Center exhibit solar-like metallicity.
It should also be noted that 
so far, no one has used Cepheid variable stars to determine
the metallicity within the young Nuclear Disc.

Some similar results have been pointed out from studies focusing
on planetary 
nebulae in the direction of the Galactic Center (see, for example,
\citealt{Cav11} 
and \citealt{Gutenkunst2008}). According to those programs, the
abundances of some 
elements in the very central  part of our Galaxy appear to be close
to the solar values. 
This is an interesting result  and it requires an independent verification
by obtaining and
analyzing the spectra of stars that definitely belong to the
Galactic Nuclear Disc.

At present, existing theoretical models focusing on the chemo-dynamical evolution of the Galaxy 
disc use modern data on the elemental yields in stars of different masses, take into 
account interstellar gas flows in the disc and halo, as well as the the dynamic influence of the Galactic bar 
on the chemical properties in the Galactic Center, etc. Therefore, elemental distributions in the 
disc obtained from observations are important criteria for verifying the reliability of those models.

For instance, \cite{Min13} performed a study of the Galaxy thin disc chemical evolution through models. According 
to their results, the metallicity of [Fe/H] in the disc gradually increases from a Galactocentric 
distance of 15 kpc to 2 kpc, reaching a plateau in 0 to 2 kpc range, with a maximum 
value of [Fe/H] of about +0.7 dex (see their Fig. 2). Similarly, \cite{Cav14} obtained a plateau in the 
radial iron distribution in the range of Galactocentric distances from 0 to 4 kpc. The maximum value [Fe/H] obtained 
by those authors is +0.4 dex.

Additionally, \cite{Kub2015} have developed models including calculations 
up to a Galactocentric distance 1 kpc. Although their [Fe/H] value at
a distance of 3--4 kpc from the Galactic Center is in good agreement
with our observational data 
from \citealt{And2016} (a small plateau in the iron abundance distribution
within this 
zone with [Fe/H] $\approx$ +0.4 dex, see their Fig. 5), their model predicts
a steady 
grows of metallicity toward the Center with [Fe/H] $\approx$ +0.7 dex at a
Galactocentric distance of 1 kpc.

Recent results presented by \cite{Toyouchi2018} show that their model predicts a metallicity [Fe/H] at the level 
of about +0.6 dex at a Galactocentric distance of 2 kpc (which is close to our 
observational result, see \citealt{And2016}). Unfortunately those authors did
not study the chemical properties of the stellar disc component in the Milky Way 
Center.
   
Recently \cite{Dekany2015} and \cite{Matsu2015} reported the photometric
discovery of several classical Cepheids which are distributed
around and behind the Galactic Center. To observe spectroscopically the
Galactic Center Cepheids is a very difficult task. Extremely strong 
light absorption makes it impossible to get high resolution spectra of
those stars in the visible. Therefore we decided to observe several targets from the list
of \cite{Matsu2016} in H spectral band in order to avoid significant
light absorption. It should be also noted that recently \cite{Inno2019}
applied IR spectra of medium resolution ($R = 3000$) of five newly discovered Cepheids 
in the direction of the Galactic Center, but their program stars are
situated at a few kpc from the center.

\section{Observations and data reduction}

Near-IR spectroscopic observations were carried out in remote observing mode
on two half-nights (May 11 and May 17, 2017), with the InfraRed Telescope
Facility (IRTF) 3-meter telescope on Maunakea.  We used the recently commissioned
cross-dispersed echelle spectrograph  iShell
(1.08--5.3 micron, R = 80000, \citealt{Rayner2016}).
The detector was binned 2$\times$2 resulting in a spectral resolution
R $\approx$ 35\,000; a 0.75$''$$\times$5$''$ slit was used for all observations.
With the H1 instrumental configuration for the spectrograph, the wavelength
coverage extended from 1.48 to 1.67 $\mu$m, a bandpass including important
spectral lines needed for the abundance analysis of our program Cepheids.
Observations were conducted under photometric skies, with a seeing
of 0.5$''$ $-$ 0.8$''$ in the visible. For each Cepheid star
(typical magnitude H $\approx$ 12), exposures were 6 $\times$ 600 seconds
(coadds). The latter strategy was selected to achieve S/N $\geq$ 30,
a minimum necessary to perform the analysis of the main spectral lines
within the H1 bandpass; the exposure times were estimated using the iShell
performance established during commissioning of the instrument by IRTF.

A reference solar spectrum (4 $\times$ 5 seconds) was also obtained by
observing the Moon. This spectrum is used to verify values for transition
oscillator strengths known from literature. A series of spectra was also obtained on 
several well-studied F and G supergiants. 

\begin{figure*}
\includegraphics[width=15cm]{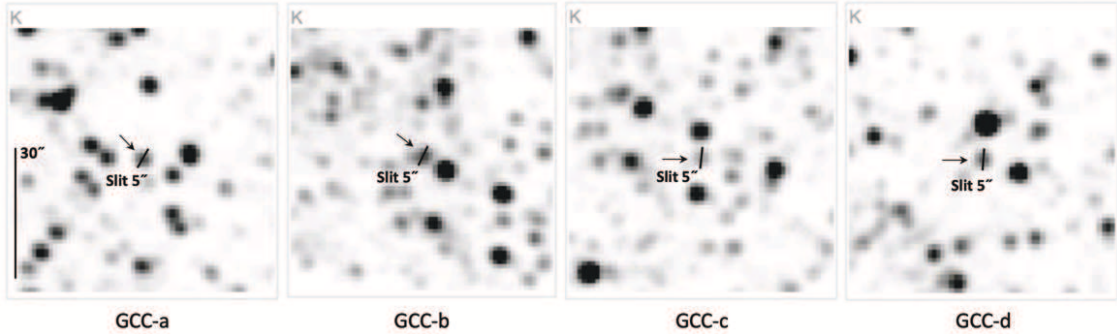}
\caption[]{Finding charts for the four inner disc Cepheids
studied for this program. The field-of-view is 60$''$$\times$ 60$''$,
similar to the field available for the iShell spectrograph guider;
sky images were extracted from the 2MASS survey in K-band.
Each individual field is centered on the coordinates for the
Cepheids as published by \cite{Matsu2016}; each target put
within the iShell slit is identified by an arrow. The field scale is displayed in the left image. 
The slit is by default at the parallactic angle. We took care
that for each angle, no other star than our target was seen in the slit.
The 5$''$ slit is displayed directly on the target, including the angle of the slit as used
during the observation. It is clear that there is no contamination
from field stars.
North is at the top, east at the left.}
\label{charts}
\end{figure*}

An observational challenge for our program was to correctly identify the right 
targets to position them within the spectrograph slit since all inner disc
Cepheids are obviously located in very crowded fields. To achieve
this, we first applied offsets relative to nearby reference stars instead of 
only relying on the accuracy of the telescope absolute pointing. These 
telescope offsets were calculated using the precise coordinates given by 
\cite{Matsu2016} for our four Cepheids.  From our experience, this technique 
brought the targeted Cepheid within a few arcseconds from the spectrograph slit. 
We then compared the guider field of view with finding charts (see Figure \ref{charts}) 
centered of the Cepheid coordinates from \cite{Matsu2016} and extracted from K-band 
images from the 2MASS survey using the Finder Chart tool at irsa.ipac.caltech.edu/applications/finderchart/. 
All Cepheid stars of our program were clearly visible in the guider; direct comparison with 
the 2MASS finding charts was possible since we also used a K-band filter in the
telescope guider (field-of-view of 42$''$). A final (manual) offset was then applied to bring 
the target Cepheid precisely within the narrow slit of the spectrograph. 
As shown also in Fig. \ref{charts}, the iShell slit is only 5$''$ in length and care was taken as
well to avoid spectral contamination by other stars within the field, at least from objects 
we could visually detect within the guider.

For wavelength calibration and for removing the telluric absorption
lines, we also observed telluric standard stars (B- and A-type
dwarfs). Spectra were combined into a single spectrum for each target. 
The telluric absorption lines of the target were subtracted using a spectrum 
of the corresponding telluric standard.

\begin{figure}
\includegraphics[width=8.5cm]{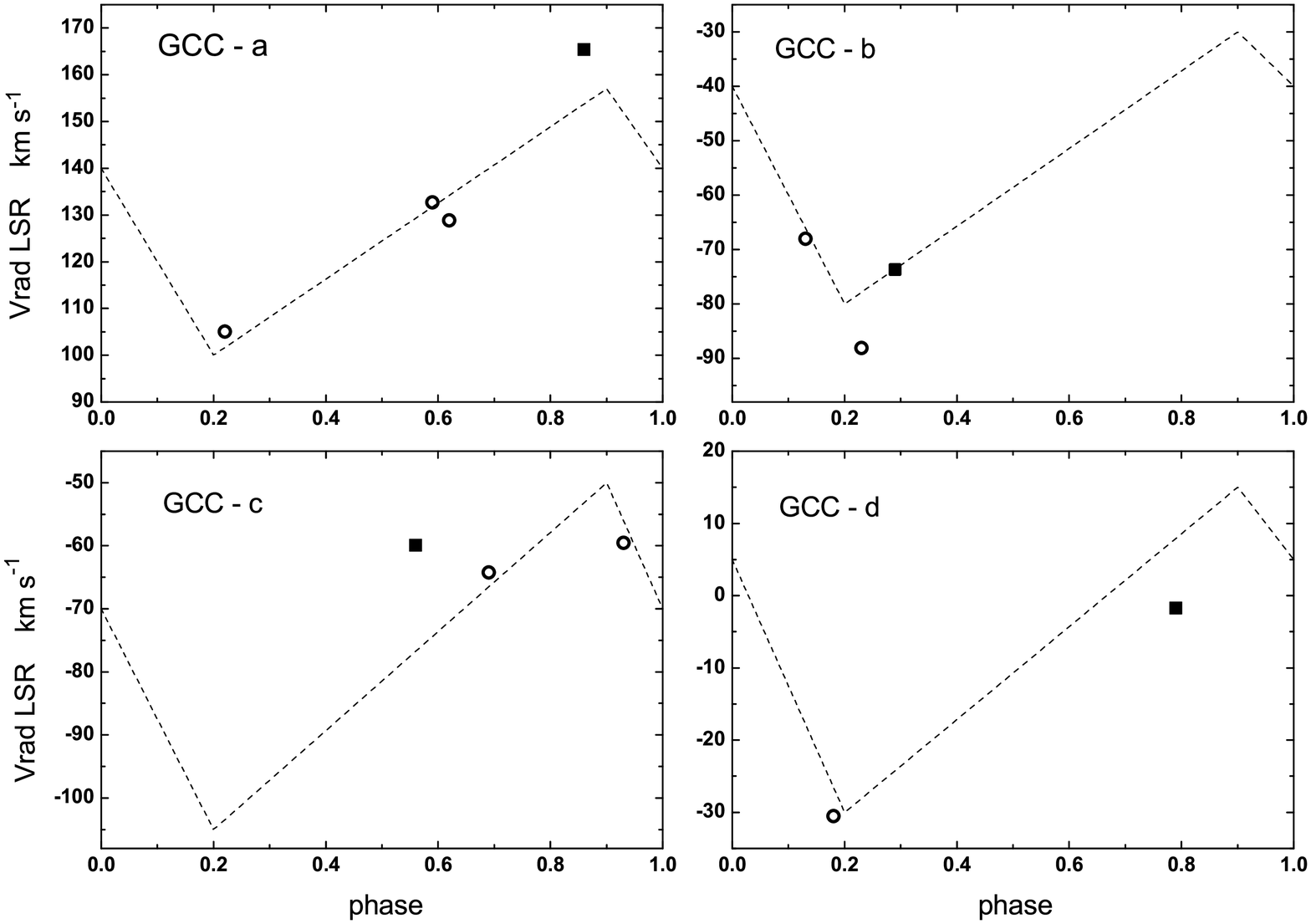}
\caption[]{Radial velocity curves of the program Cepheids. $Open~circles$ indicate data 
from the paper of \cite{Matsu2015}, $black~squares$ represent our data. $Dashed~lines$ 
schematically show radial velocity curves of Cepheids according to the data
from \citealp{Matsu2015} (see their Fig.6). Deviations of our points from the
curve data may be caused by evolutionary changes of periods.}
\label{Vrad}
\end{figure}

As an additional check that we indeed observed the same Cepheids
as sampled by  \cite{Matsu2015}, we measured the radial velocities of our targets
using telluric lines as the wavelength reference frame. The results are shown in Table \ref{Obs}. 
The velocities obtained were then transformed into barycentric velocities, 
\Vbary, and velocities relative to the local standard of rest (LSR), \VLSR, 
assuming the standard solar motion  \citep[][see Table \ref{Obs}]{Reid2009}.
All radial velocities show good fit with the radial velocity curves given in
\citealp{Matsu2015} (see our Fig. \ref{Vrad} and their Fig. 6).
Pulsational phases were calculated according to the data of \cite{Matsu2015}.
To control our results, we also determined the radial velocities of the F and G supergiants observed for our program.
Our results are in a good agreement with the literature data (see Table \ref{Comp}). 
Note that HD~182296 is a spectroscopic binary star.
Thus, we are confident that our faint targets have been correctly identified.

The details of observations are provided in Table \ref{Obs}.

\begin{table*}
\small
\caption{Journal of observations.}
\label{Obs}
\begin{tabular}{lccccccccccc}
\hline
Cepheid  & N & R.A          & Dec           & Date       & JD(start)& Phase & Integ. Time     & S/N &\Vbary &  \VLSR & Airmass  \\
          &   &      J2000.0 &      J2000.0  &       UTC  & 2457000+ &       &      s          &     & \kms  & \kms   &          \\
\hline
GCC-a& 13  & 17:46:06.0    & $-$28:46:55.1  & May 18, 2017  & 892.0150 &  0.86 & 7 $\times$ 600& 30  & 155.0 & 165.4  & 1.52  \\  
GCC-b& 12  & 17:45:32.3    & $-$29:02:55.3  & May 18, 2017  & 891.9552 &  0.29 & 6 $\times$ 600& 35  &--84.0 &--73.7  & 1.60  \\  
GCC-c& 11  & 17;45:30.9    & $-$29:03:10.6  & May 12, 2017  & 886.0177 &  0.56 & 6 $\times$ 600& 33  &--70.2 &--59.9  & 1.51  \\  
GCC-d& 10  & 17:44:56.9    & $-$29:13:33.8  & May 12, 2017  & 885.9623 &  0.79 & 6 $\times$ 600& 37  &--11.9 & --1.7  & 1.61  \\  
           &               &                &               &          &       &               &     &       &        &       \\
HD 172594& & 18:41:42.5    & $-$14 33 51.3  & May 12, 2017  & 886.0595 &  --   & 2 $\times$ 20 & 100?& --2.8 &  11.2  & 1.21  \\
HD 179784& & 19:13:15.4    &  +15 02 08.3   & May 18, 2017  & 892.0755 &  --   & 3 $\times$ 10 & 100?&--21.2 & --2.6  & 1.00  \\
HD 182296& & 19:23:38.7    &  +08 39 36.0   & May 18, 2017  & 892.0781 &  --   & 4 $\times$ 10 & 100 &--12.0 &   5.6  & 1.02  \\
\hline
\end{tabular}

Remarks:
The name of Cepheid is from \cite{Matsu2015}.
The number N is given according to the designation of \cite{Matsu2016}.
Phases were calculated according to the data of \cite{Matsu2015}.
\end{table*}

\section{Spectroscopic analysis}

\begin{figure}
\includegraphics[width=8.5cm]{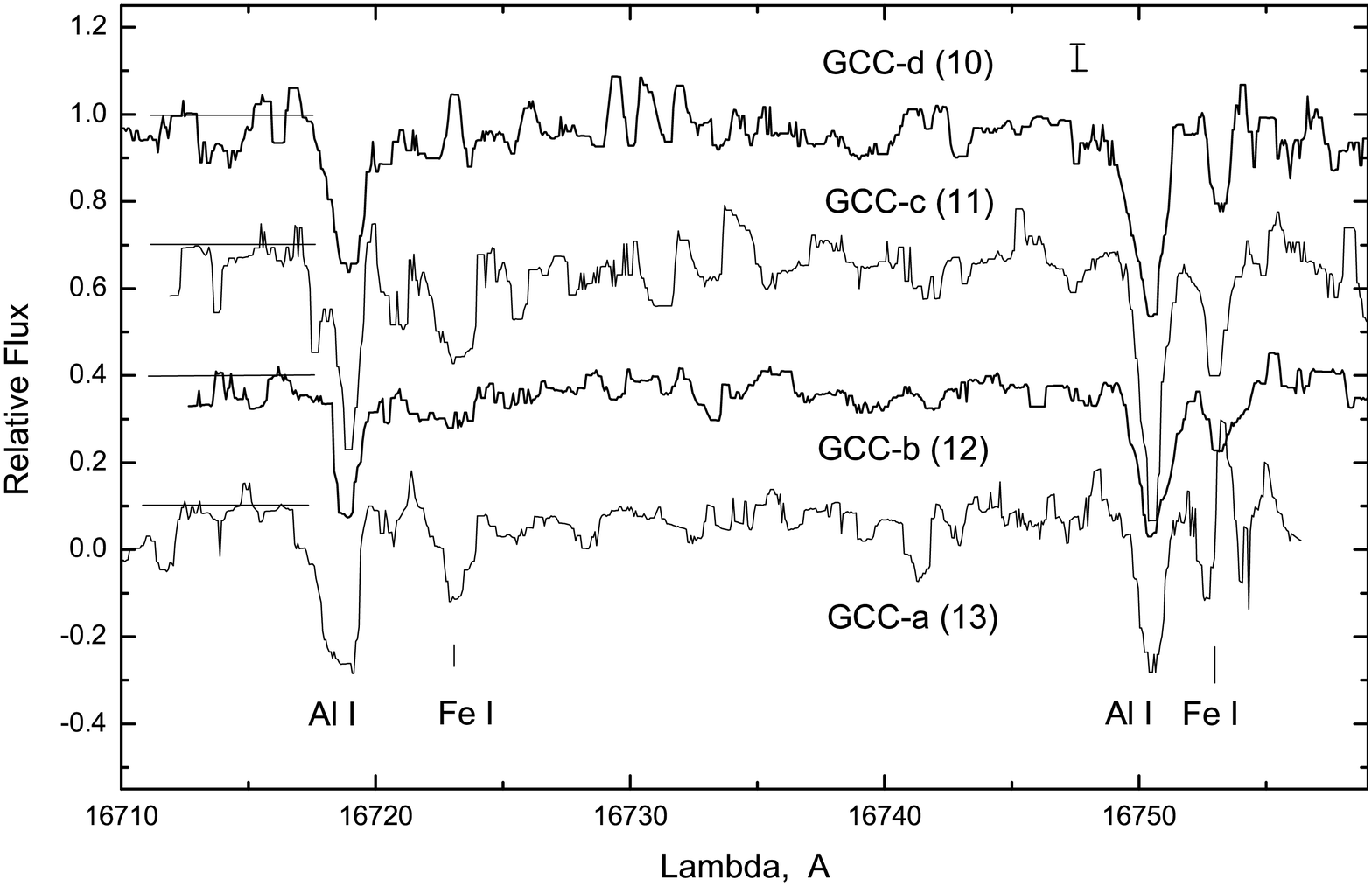}
\caption[]{Fragments of the spectra of program Cepheids. 
A portion of the continuum for each spectrum is shown on the left by a
$continuous~line$.}
\label{sp}
\end{figure}

In order to normalize the individual spectra to the local continuum, 
to identify the lines of different chemical elements, and to measure 
the equivalent widths (EW) of the absorption lines, we used the DECH~30 
software package\footnote{http://www.gazinur.com/DECH-software.html}.
Fragments of the spectra of our program Cepheids are shown in Fig. \ref{sp}.

As we can see from Fig. \ref{sp}, the spectra are rather noisy, but some lines
can be measured with a reasonable precision.

The effective temperatures (\Teff) of our program F--G supergiants have been
estimated using the calibrating IR line depth ratios, as explained in \cite {Fukue15}. 
In the Table \ref{Comp} we give the resulting temperature values determined using 
IR and visual spectra. The values of \Teff\ for visual spectra are based on the use 
of  \cite{Kov2007} calibrating line depth ratio method. As can be seen, independent 
estimates are in fairly good agreement.

\begin{table*}
\caption{Comparison of the temperature and radial velocity values for
supergiants determined from visual and H-band spectra for three supergiants.}
\label{Comp}
\begin{tabular}{cccccccrrcc}
\hline
  & &\multicolumn{3}{c}{\Teff (visual)} &
\multicolumn{3}{c}{\Teff (H-band)}&\multicolumn{3}{c}{Radial velocity}\\
  & &\multicolumn{3}{c}{ \citep{Kov2007}} &
\multicolumn{3}{c}{(this paper)}& (H-band)& (Literature) \\
\cline{2-10}
HD      & Sp&\Teff  &$\sigma_{mean}$ &N&$T_{\rm eff}$&$\sigma_{mean}$&N& \Vr & \Vr & \\
        &   &   K   &    K   & &   K    &   K  & & \kms & \kms   \\
\hline
 172594 &F2Ib& --   &  --     & --  &   --&    --    &  --    & --2.8 & --2.41             \\
 179784 &G5Ib&  4956& $\pm$42 & 55  & 5054& $\pm$123 &  8     &--21.2 &--21.67             \\
 182296 &G3Ib&  5072& $\pm$50 & 77  & 5059& $\pm$88  &  3     &--12.0 &--9.35...--14.0 SB  \\
\hline
\end{tabular}

Remark: N is a number of used temperature calibrations.

Typical literature \Vr\, values can be found e.g. in SIMBAD.
\end{table*}

The effective temperatures of our program Cepheids have also been
estimated using the calibrating line depth ratios from \cite {Fukue15} (see Table \ref{Par}).

For bright supergiant stars, the surface gravity is estimated from the ionizational balance for iron using
the visual spectrum; this value is then adopted for the near-IR spectroscopic analysis (measurable Fe{\sc ii}
lines are not available in the H-band region).
Since we had not enough Fe{\sc ii} lines for Cepheids, we were not able to estimate surface gravity 
using the typical condition of ionization balance. In this case we roughly evaluated \logg~with the
help of approximate relation between surface gravity of Cepheids and their pulsational period 
\citep[see Fig. 5 in][]{And05}.

The microturbulent velocity \Vt\ for our program stars was found by avoiding any dependence 
between the iron abundance as produced by individual Fe{\sc i} lines and their equivalent widths. 

The resulting atmosphere parameters and other information for the  studied stars are listed in Table \ref{Par}.

\begin{table*}
\small
\caption{Physical parameters of the investigated Cepheids.}
\label{Par}
\begin{tabular}{rcccccccccc}
\hline
Cepheid & N&  P   &JD, 2457000+&$<$H$>$ &\Teff  &\logg & \Vt &  \Rg   \\
        &  &   day   & (start) &  mag   &  K    &      & \kms&   kpc  \\
\hline
GCC-a & 13 &   23.52 & 892.0150 & 12.02 &  4850 & 1.0  & 3.5 &   $<$0.2    \\
GCC-b & 12 &   19.96 & 891.9552 & 11.96 &  5050 & 1.2  & 3.0 &   $<$0.2    \\
GCC-c & 11 &   22.75 & 886.0177 & 12.39 &  5000 & 1.2  & 3.0 &   $<$0.2    \\
GCC-d & 10 &   18.87 & 885.9623 & 12.14 &  5580 & 1.4  & 3.5 &   $<$0.2    \\
\hline
\end{tabular}

Remark:
The number N is given according to \cite{Matsu2016}.
\end{table*}

Approximated accuracy of our parameter determination is estimated to be: 

$\Delta$\Teff = $\pm200~K$, $\Delta$\logg = $\pm0.3$ dex,

$\Delta$\Vt = $\pm0.5$ \kms, $\Delta$[Fe/H] = $\pm0.2$ dex.

\section{Abundance analysis}

The local thermodynamical equilibrium elemental abundances in our program stars were 
calculated using the WIDTH9 code and ATLAS12 atmosphere models. For that purpose, we used 
newly evaluated astrophysical oscillator strengths and damping parameters adopted by 
the Apache Point Observatory Galactic Evolution Experiment (APOGEE, \citealt{Shet15}, 
their Table 7).

Abundances in two supergiants were derived  in order to check the reliability
of our H-band spectroscopic analysis. For these stars we compared abundances derived 
from H-band spectra with abundances derived from the visual spectra. Results for H-band 
are presented in Table \ref{Super} for one star, HD~179784. 

Resulting abundances for our program Cepheids are given in Table \ref{Abund}.

\begin{table}
\caption[]{Comparison of abundances in HD 179784 derived from visual and H-band 
spectra. Stellar atmosphere parameters are presented in the form \Teff/\logg/\Vt}
\label{Super}
\begin{tabular}{lrrrrrrr}
\hline
\hline
\noalign{\smallskip}
\multicolumn{1}{c}{}&\multicolumn{3}{c}{HD179784 (Visual)}&
\multicolumn{3}{c}{HD179784 (H-band)}\\
\multicolumn{1}{c}{}&\multicolumn{3}{c}{4956/1.8/2.3}&
\multicolumn{3}{c}{5054/1.8/2.2}\\
\noalign{\smallskip}
\hline
 Ion &[M/H]&$\sigma$&N&[M/H]&$\sigma$&N\\
\hline
 6.00 &  --0.18 & 0.24 &    3 &  --0.40 & 0.18 &    6   \\
 7.00 &    --   & --   &   -- &    0.37 & --   &    1   \\
 8.00 &    0.16 & --   &    1 &    --   & --   &   --   \\
11.00 &    0.40 & 0.02 &    2 &    0.38 & 0.16 &    2   \\
12.00 &  --0.11 & --   &    1 &    0.21 & 0.04 &    2   \\
13.00 &    0.18 & 0.13 &    2 &    --   & --   &   --   \\
14.00 &    0.07 & 0.08 &   10 &  --0.03 & 0.14 &   13   \\
15.00 &    --   & --   &   -- &    0.36 & 0.19 &    3   \\
16.00 &    0.35 & --   &    1 &    0.04 & 0.12 &    8   \\
20.00 &    0.02 & 0.08 &    3 &    0.24 & 0.07 &    3   \\
21.01 &    0.05 & 0.11 &    5 &    --   & --   &   --   \\
22.00 &  --0.21 & 0.13 &   14 &    0.12 & 0.10 &    6   \\
22.01 &  --0.16 & 0.13 &    3 &    --   & --   &   --   \\
23.00 &  --0.13 & 0.12 &    7 &    0.02 & 0.53 &    5   \\
23.01 &    0.10 & 0.13 &    4 &    --   & --   &   --   \\
24.00 &  --0.17 & 0.10 &   11 &    0.02 & 0.02 &    3   \\
24.01 &    0.22 & 0.10 &    3 &    --   & --   &   --   \\
25.00 &    0.10 & 0.11 &    5 &  --0.04 & 0.40 &    6   \\
26.00 &    0.00 & 0.10 &   82 &    0.12 & 0.15 &  156   \\
26.01 &    0.02 & 0.12 &   14 &    --   & --   &   --   \\
27.00 &  --0.03 & 0.14 &    8 &    0.16 & 0.46 &    6   \\
28.00 &  --0.03 & 0.10 &   36 &    0.04 & 0.13 &    8   \\
29.00 &  --0.41 & 0.00 &    1 &  --0.11 & 0.33 &    2   \\
39.01 &    0.22 & 0.07 &    4 &    0.36 & --   &    1   \\
40.01 &  --0.03 & 0.02 &    2 &    --   & --   &   --   \\
57.01 &    0.17 & --   &    1 &    --   & --   &   --   \\
58.01 &    0.01 & 0.04 &    4 &    --   & --   &   --   \\
59.01 &  --0.06 & 0.05 &    2 &    --   & --   &   --   \\
60.01 &    0.12 & 0.07 &    5 &    --   & --   &   --   \\
63.01 &    0.21 & 0.01 &    2 &    --   & --   &   --   \\
64.01 &    0.38 & --   &    1 &    --   & --   &   --   \\
\hline
\end{tabular}
\end{table}

\begin{table*}
\caption[]{Abundances of individual elements for Galacic Center Cepheids.}
\label{Abund}
\begin{tabular}{lrrrrrrrrrrrrr}
\hline
\hline
\noalign{\smallskip}
\multicolumn{1}{c}{}&\multicolumn{3}{c}{GCC-a (13)}&
\multicolumn{3}{c}{GCC-b (12)}&\multicolumn{3}{c}{GCC-c (11)}&
\multicolumn{3}{c}{GCC-d (10)}\\
\noalign{\smallskip}
\hline
 Ion &[M/H]&$\sigma$&N&[M/H]&$\sigma$&N&[M/H]&$\sigma$&N&[M/H]&$\sigma$&N\\
\hline
 6.00 &--0.21&  0.21 &  4 &--0.30&  0.16&   5&--0.40 & 0.22 & 3& --0.25 & 0.11&    5\\
11.00 &  0.61&  0.09 &  2 &  0.46&  0.09&   2&  0.84 &  --  & 1&   0.62 & 0.04&    2\\
12.00 &  0.38&  0.17 &  2 &--0.09&  0.30&   2&  0.11 &  --  & 1&   0.03 & 0.05&    2\\
14.00 &  0.04&  0.18 &  8 &  0.01&  0.21&  13&  0.41 & 0.28 & 5& --0.14 & 0.03&    3\\
16.00 &--0.19&  0.22 &  2 &--0.44&  0.10&   6&--0.05 & 0.20 & 6&   0.09 & 0.12&    6\\
20.00 &--0.02&  0.30 &  2 &--0.36&  0.18&   3&  --   & --   &  & --0.07 & 0.43&    3\\
22.00 &  0.37&  0.33 &  4 &  --  &   -- & -- &  0.01 & 0.39 & 2&   0.27 & 0.11&    2\\
23.00 &--0.42&   --  &  1 &--0.04&   -- &   1&  --   & --   &  &   0.39 & 0.25&    2\\
24.00 &  0.27&  0.36 &  2 &--0.23&  0.11&   3&  0.44 & --   & 1&   --   & --  &   --\\
26.00 &--0.04&  0.13 & 57 &--0.01&  0.22& 106&  0.16 & 0.21 &49&   0.04 & 0.18&  122\\
27.00 &--0.24&  0.10 &  2 &  --  &  --  &  --&  0.38 & --   & 1& --0.08 & 0.03&    2\\
28.00 &  0.13&  0.22 &  8 &  0.13&  0.20&   7&--0.05 & 0.13 & 2&   0.18 & 0.33&    4\\
39.01 &  --  &  --   & -- &  0.11&  --  &   1&--0.48 & --   & 1& --0.01 & -- &    1\\
\hline
\end{tabular}
 \end{table*}

\begin{figure}
\includegraphics[width=8.5cm]{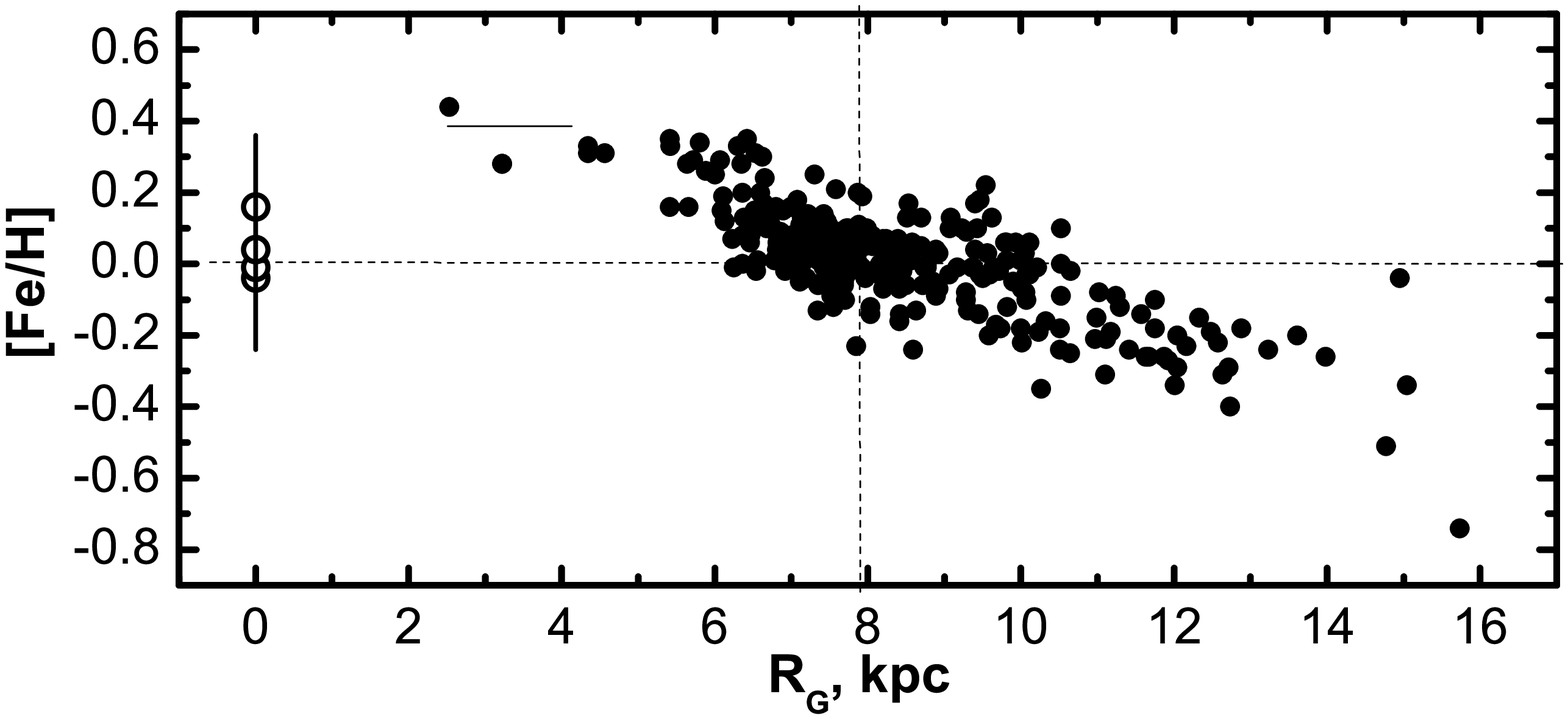}
 \caption[]{[Fe/H] vs \Rg. $Filled~circles$ -- compilation of iron abundance determinations in Galactic Cepheids from 
our papers published from 2002 to 2016 \citep[see][for references]{Mar2015, And2016}.
The iron abundance in the Galactic Nuclear Center Cepheids is from this paper
(the \Fe\ values are indicated by the $open~circle$ symbols with error bars).
$Two~points$ at approximately 2 and 3 kpc correspond to
ASAS181024-2049.6 and SU Sct positions, respectively. Approximate
position of a plateau-like region (if it really exists )
is schematically shown  by a $thin~continuous~line$. The position
of the Sun is at the intersection of the $dashed~lines$.}
\label{FeRg}
\end{figure}

\section{Results and discussion}

The steady decrease of metallicity with Galactic radius as observed for Cepheids can be interpreted
in a simple way. Excluding the possibility of a large radial excursion of the
Cepheids, since these stars are very young, 
their metallicity reflects the local abundance of the interstellar medium
(ISM), at their current Galactic radius. The enrichment of
the ISM is mostly due to the explosion of supernovae \citep{Matteucci2009}.
Since the Initial Mass Function seems to be universal, the formation
rate of stars of all masses in just in proportions given by the IMF. 
A region were the density of stars is large is a region where the star 
formation rate and the duration of the star forming process were large. 
Therefore,  we expect that the metallicity at a given radius of the disc 
should be proportional to the stellar density.
Thus, it is important to establish what the stellar density is in the
thin disc as a function of radius. Many authors consider that the disc
density is exponential \citep{Robin2003} according to a law which extends
(or not) up to the center. A disc profile proposed by \cite{Kormendy1977},
mostly based on observations of external galaxies, seems close to reality.
The law for the surface density is
$\Sigma =  \Sigma_0 \exp( -r/\alpha - (\beta/r)^n)$, where $\alpha$
is the scale length of the disc and $\beta$ the "size " of the hole;
$n$ is taken equal to 2. This law produces a decrease of $\Sigma$ for values 
of $r$ smaller than $\beta$. The metallicity data is a logarithmic scale, so that 
we should compare it with  $\log(\Sigma)$. If we adopt $\beta$ of the order of 
2 kpc, following  \citet{LL2000}, this disc profile predicts a maximum metallicity
around $r = 2$ kpc, in satisfactory agreement with our Cepheids' data.
 
However, the central region of the Galaxy is far from being axi-symmetric,
due to the presence of the bar. A recent sketch of the bar, with its extension
and orientation in a direction not far from that of the Sun (about $20^o$),
is presented by \citet{Michtchenko2018} (their Figure 15).
 In the plane of the Galaxy and for radii $<$ 3 kpc (in the direction of
the axis of the bar) the bar dominates the stellar population.
There is no star formation in an extended region. The masers with
precise VLBI distances measured by \cite{Reid2014}, and which are associated
with massive stars, are plotted in that figure. The presence of masers
signals the regions where very young  stars are present. We can see that
there are masers all around the bar, but not inside it.
Observations of external barred galaxies confirm that most bars are
of yellowish color, which indicates that they are mainly constituted of old
stellar population. Except maybe for some very late-type, low-mass or weakly 
interacting barred spirals (see for instance \citealp{MF97}),
there are no O--B stars in the bar, except at their
extremities and at their very center. This also explains the absence of
numerous Cepheids in the Galactic radius range 1--3 kpc.
 
The region very close to the center (\Rg$<$1 kpc) is quite different from its
surroundings. It contains at least one maser with distance determined
by VLBI measurements, at least 4 Cepheids (the ones studied in this work)
and a major molecular cloud, Sagittarius B2. The physical conditions in
this area are possibly different from those prevailing in the Solar
neighborhood. There is possibly an inflow of matter with a high metallicity
coming from internal regions to the disc, flowing across the bar, and possibly 
an inflow of low metallicity gas from the spheroidal bulge. It is beyond
the scope of the present work to develop a specific chemical evolution model
for the Galactic Center.

In Fig. \ref{FeRg} we show the iron abundance distribution from the Galactic Cepheids 
based on our previous studies (see references in \citealt{Mar2015} and \citealt{And2016}) 
and the central value obtained in the present study. We were especially interested in the iron 
content, since this element has the largest number of lines in the spectra studied, and its 
content was found with a sufficiently high accuracy. We also have the largest statistics 
on abundance distribution in the disc for iron. In our previous papers we reported on the elemental 
distribution in the range of Galactocentric distances extending from about 2 kpc to 16 kpc.

The Cepheids of our present program cover a similar pulsation period range as the stars included 
in our previous samples; thus they share a similar age range. Following the period-age relation for 
classical Cepheids as proposed by \citet{Bono2005}, the age range of the Cepheids in our sample is 
roughly 20--70 Myr. So, these are young stars, and they should be situated very near their birthplaces.

The average iron abundance  at the Galactic Center from literature is presented in \cite{Mar2015}
and \cite{And2016}. That value was mostly determined from the stars covering a larger age range 
(some ofthem are members of the Arches/Quintuplet clusters with age of
about $\approx$ 3--9 Myr, while there are the field stars with age
of $\approx$ 1Gyr. Our data on four young classical Cepheids apparently suggest
that the metallicity at the very center of our Galaxy disc is approximately solar.
This is the main result of this paper. Looking at Fig. \ref{FeRg}
one can note that the metallicity gradually increases from the outer part of the
thin disc reaching the maximum value of about +0.4 dex at Galactocentric distance
in the range from 2 to 4 kpc, and then decreases to about the solar value in the Galaxy
Center. In the range 2--4 kpc,  we yet cannot confirm or disprove some kind of the 
plateau-like structure in the metallicity distribution as it was first supposed in
\cite{And2016}, even with our additional data. More observations will be needed. Independent 
arguments supporting our result can be found in a recent paper by \cite {Bovy2019}, 
which analyzed the APOGEE observational data. These authors clearly found a zone of increased 
metallicity at a Galactocentric distance of about 4 kpc, while the metallicity at the Center was 
similar to that measured at the solar Galactocentric distance (see their Fig. 4).

\begin{figure}
\includegraphics[width=8.5cm]{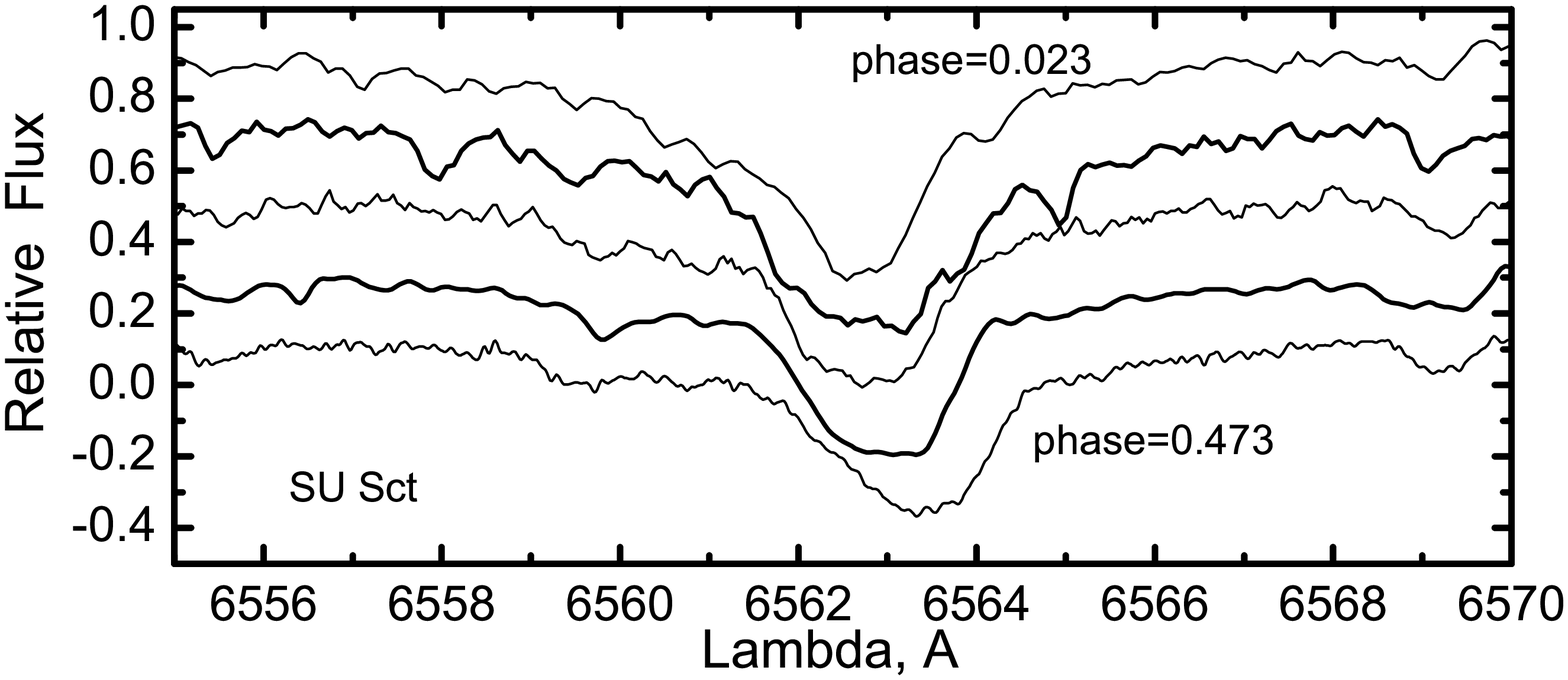}
\caption[]{Variation of the H$\alpha$ profile in SU Sct spectra with
the pulsation phase. None of the spectra (including the spectrum that was observed at the maximum light) 
shows any emission feature in this line. Three spectra were collected by
Prof. George Wallerstein using the 
facilities of Apache Point Telescope (priv. comm.) and, the other two
were investigated by \cite{Mar2015}.}
\label{Halpha}
\end{figure}

How reliable is our conclusion regarding the maximum abundance at 2--4 kpc? The answer significantly 
depends on the position of two stars, namely SU Sct and ASAS 181024-2049.6, which were
previously analyzed in \cite{Mar2015} and \cite{And2016}. In those papers we proceeded
from the assumption that both are classical Cepheids.
\cite{Jayasinghe2018} performed the All-Sky Authomated Survey for Supernovae
and, as a by-product, they discovered a large number of new variable stars.
In particular they found that SU Sct could be a W Vir star, i.e. Cepheid of
type II. If this classification is correct, then the distance used in
our previous program may be wrong. Nevertheless, it should be noted
that the above authors carried 
out an automatic classification of a large sample of stars without
a careful investigation of the individual objects.
Therefore, it cannot be excluded that in the case of SU Sct the
classification presented may be erroneous.
Moreover, type II Cepheids are known to  show a strong emission in H$\alpha$ near maximum, 
whereas classical Cepheids, as a rule, do not show it. The star of our program SU Sct does not show 
emission in H$\alpha$ (we have five spectra of this Cepheid in our disposal, see fragments of
those spectra in Fig.\ref{Halpha}). In addition, type II Cepheids have a metallicity below zero,
while SU Sct shows a metallicity  of about +0.3 dex. So it is highly probable that SU Sct is a classic Cepheid.
ASAS 181024-2049.6 (three spectra obtained in different phases) also does not show any H$\alpha$ emission,
and it also likely a classic Cepheid. Moreover, both these stars are situated very close to the Galactic disc
($b$ is about --6$^{\circ}$ and --1$^{\circ}$ respectively), which is not typical for type II Cepheids.

\section{Conclusion}

For the first time we derived elemental abundance in fairly young disc stars -- Cepheids, 
which are located in the very center of our Galaxy. We used high-resolution near-IR spectra of 
four stars observed with the iShell spectrograph attached to the NASA InfraRed Telescope Facility. 
Our LTE analysis showed that these Cepheids have metallicities close to solar values,
a new result for the very central part of the Galactic Nuclear Disc.

\section{Acknowledgments}

We acknowledge the excellent support team at NASA IRTF and for the iShell
echelle spectrograph. 
The authors wish to recognize and acknowledge the
very significant role that the summit of Maunakea has always had within
the indigenous Hawaiian community. We are most grateful to have had the
opportunity to conduct observations from this mountain over the years.
WJM thanks CNPq (Process 302556/2015-0) and FAPESP (Process 2010/18835-3
and 2018/04562-7). Authors are also thankful to the anonymous referee for his/her 
valuable comments which have considerably improved our paper.

\label{lastpage}

\bsp

\end{document}